\documentclass[lettersize,journal]{IEEEtran}
\usepackage{amsmath,amsfonts}
\usepackage{algorithmic}
\usepackage{array}
\usepackage[caption=false,font=normalsize,labelfont=sf,textfont=sf]{subfig}
\usepackage{textcomp}
\usepackage{stfloats}
\usepackage{url}
\usepackage{verbatim}
\usepackage{graphicx}
\def\BibTeX{{\rm B\kern-.05em{\sc i\kern-.025em b}\kern-.08em
    T\kern-.1667em\lower.7ex\hbox{E}\kern-.125emX}}
\usepackage{balance}

\begin{document}
\title{From 6G Scenarios and Requirements to Design Drivers: Insights from 3GPP Release 20}
\author{Victor Monzon Baeza,~\IEEEmembership{Senior Member,~IEEE}, and Symeon Chatzinotas, ~\IEEEmembership{Fellow,~IEEE}  
\thanks{Manuscript received XXX, XX, 2026; revised XXX, XX, 2026.}
\thanks{V.M. Baeza is with Escuela Superior de Ingeniería y Tecnología (ESIT), Universidad Internacional de La Rioja (UNIR), Logroño, 26006 La Rioja, Spain. S. Chatzinotas is with Interdisciplinary Centre for Security, Reliability and Trust (SnT), University of Luxembourg, Luxembourg. \textit{Corresponding author: V. M. Baeza (victor.monzon@unir.net).}
}}

\markboth{Journal of \LaTeX\ Class Files,~Vol.~XX, No.~X, March~2026}%
{}

\maketitle

\begin{abstract}
The definition of sixth-generation (6G) systems is being shaped by early standardization efforts, including the 3GPP TR 38.914 (Release 20) study on scenarios and requirements. This study introduces a comprehensive set of deployment environments, service classes, and performance targets that will guide the evolution toward IMT-2030.
This article provides a design-oriented interpretation of these definitions, bridging the gap between standardized scenarios and system design. We first organize 6G deployment scenarios and emerging services into a unified framework. We then identify key design drivers derived from the 3GPP requirements, including terrestrial–non-terrestrial integration, GNSS-free operation, AI-native networking, and joint communication and sensing.
Finally, we discuss the implications of these drivers on 6G architecture and highlight open challenges for future standardization and research.
\end{abstract}
 
\begin{IEEEkeywords}
3GPP, 6G, AI, IMT-2030, NTN, Design Drivers, TR 38.914, Release 20
\end{IEEEkeywords}

\section{Introduction}
\label{sec1}

The evolution toward sixth-generation (6G) wireless systems is expected to redefine the role of communication networks as enablers of tightly integrated digital and physical worlds \cite{saad2019vision,strinati20216g,akyildiz2020moving}. Beyond incremental improvements in throughput or latency, 6G is envisioned to support new paradigms such as integrated sensing and communication \cite{11359163}, native artificial intelligence (AI) functionalities \cite{11455199}, and seamless connectivity across terrestrial and non-terrestrial domains \cite{10570308}. These ambitions require a fundamental rethinking of system design, where architectural choices are increasingly driven by heterogeneous deployment conditions and diverse service demands.

In this context, early standardization activities play a critical role in shaping the technological direction of 6G. In particular, the 3GPP Technical Report (TR) 38.914, which studies scenarios and requirements for Release 20, provides one of the first comprehensive attempts to capture the expected operational environments, service categories, and performance objectives of future 6G systems \cite{3gpp38914}. The study spans a wide range of deployment scenarios, including dense urban areas, industrial environments, high-mobility use cases, and non-terrestrial networks (NTN), and associates them with emerging services such as immersive communications, massive machine-type communications, sensing, and AI-driven applications.

While these definitions offer a valuable foundation, their implications for system design are not explicitly articulated. The study primarily presents scenarios, requirements, and key performance indicators as separate elements, leaving open the question of how these components jointly influence the design of future radio access technologies and network architectures. As a result, there is a lack of a structured interpretation that connects standardization inputs with concrete design directions, which is essential for both industry practitioners and researchers navigating the early stages of 6G development.

This article addresses this gap by providing a design-oriented framework that links 6G scenarios and requirements to the underlying system design drivers. Rather than summarizing the standard, the goal is to extract and organize the most relevant elements into a coherent perspective that highlights their combined impact on 6G system design. First, we introduce a unified taxonomy of deployment scenarios that captures their fundamental characteristics in terms of density, coverage, mobility, and infrastructure. Next, we analyze the emerging service classes and their associated capability demands. Building on this foundation, we identify a set of key design drivers that are implicitly defined across the 3GPP study, including the convergence of terrestrial networks and NTN, support for operation without reliance on global navigation satellite systems (GNSS), native integration of AI within the network, and the joint provision of communication and sensing functionalities.

\begin{figure}
    \centering
    \includegraphics[width=1\linewidth]{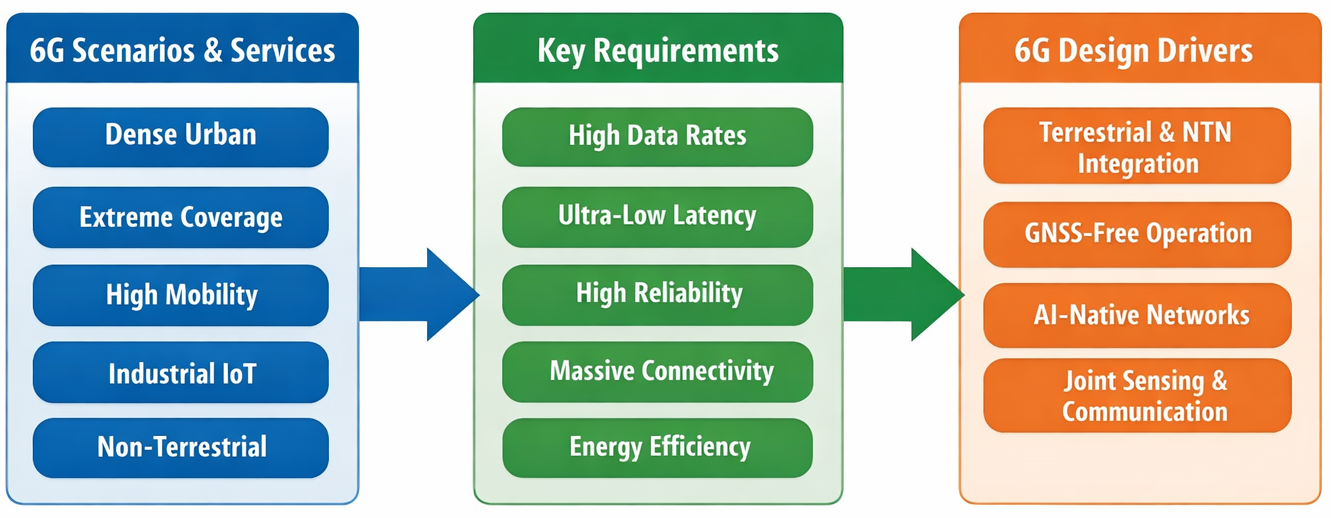}
    \caption{Mapping of 6G scenarios and service classes to system requirements and resulting design drivers, based on 3GPP TR 38.914 Release 20.}
    \label{fig:framework}
\end{figure}

Finally, we discuss how these design drivers influence architectural choices, radio design, and system operation, and outline open challenges that remain to be addressed in future standardization efforts. By translating early 6G standardization inputs into a structured design perspective, this work aims to support the development of more coherent and scalable 6G solutions across research, industry, and standardization communities.
This relationship is illustrated in Fig.~\ref{fig:framework}, which shows how diverse deployment scenarios and emerging services translate into key performance requirements, ultimately shaping the fundamental design drivers of 6G systems.

\section{Overview of the 3GPP TR 38.914 (Rel-20)}
\label{sec:over}

The 3GPP TR 38.914 Release 20 study on scenarios and requirements represents one of the first coordinated efforts to define the foundations of 6G systems from a radio access network (RAN) perspective \cite{3gpp38914}. As a study item, it does not specify concrete technical solutions but instead establishes a structured baseline to guide subsequent standardization phases and contribute to the definition of IMT-2030 \cite{itu2030} performance requirements.

The study adopts a scenario-driven methodology, where a diverse set of deployment environments is first identified and then associated with corresponding service demands and system capabilities. These scenarios span a wide range of operational conditions, including dense urban deployments, rural and extreme coverage areas, high-mobility use cases, industrial environments, and NTNs. Each scenario is characterized by attributes such as user density, mobility, coverage range, and spectrum usage, reflecting the heterogeneity expected in future 6G systems.

In parallel, the study defines a set of service classes that extend beyond traditional mobile broadband. These include immersive communication, massive machine-type communication, integrated sensing, AI-enabled services, and mission-critical applications. Rather than treating these services independently, the study implicitly highlights their coexistence and interaction within a single network infrastructure, suggesting a more unified and flexible system design.

Based on the identified scenarios and service classes, the study outlines a set of key system requirements that characterize future 6G systems. These include extreme performance dimensions such as very high data rates, ultra-low latency, and high reliability, as well as support for massive connectivity, ubiquitous coverage through terrestrial and non-terrestrial integration, and high-mobility scenarios. In addition, new requirement dimensions emerge, including native integration of AI, joint communication and sensing capabilities, and operation without reliance on GNSS. Energy efficiency, sustainability, and architectural scalability are also emphasized as fundamental constraints. Rather than defining fixed numerical targets, these requirements are expressed as qualitative dimensions that guide the design of future 6G systems. This shift reflects two key considerations. First, the diverse and often conflicting nature of 6G requirements makes it impractical to simultaneously achieve all performance targets, leading to inherent trade-offs across system dimensions. Second, previous generations have shown that overly rigid quantitative targets can be difficult to achieve across heterogeneous deployment scenarios, motivating a more flexible, design-oriented definition of requirements.

While the 3GPP TR 38.914 study provides a comprehensive set of inputs, its structure remains primarily descriptive. The use of qualitative dimensions reflects the need to accommodate trade-offs and heterogeneous deployment conditions, but it leaves open how these elements should be jointly interpreted from a system design perspective. Moreover, the absence of strict quantitative targets motivates a design-oriented approach focused on identifying underlying system drivers rather than specific performance values. This perspective aims to provide a coherent and scalable interpretation of early 6G standardization inputs.

\section{A Unified Taxonomy of 6G Deployment Scenarios}
\label{sec:tax}

The 3GPP TR 38.914 (Rel-20) study defines a wide range of deployment scenarios, each characterized by specific environmental and operational conditions. However, these scenarios are presented as individual cases, which makes it difficult to extract general design insights. To address this limitation, we propose a unified taxonomy that groups scenarios according to their fundamental system characteristics, enabling a more structured interpretation of their impact on 6G design.

\begin{figure}
    \centering
    \includegraphics[width=1\linewidth]{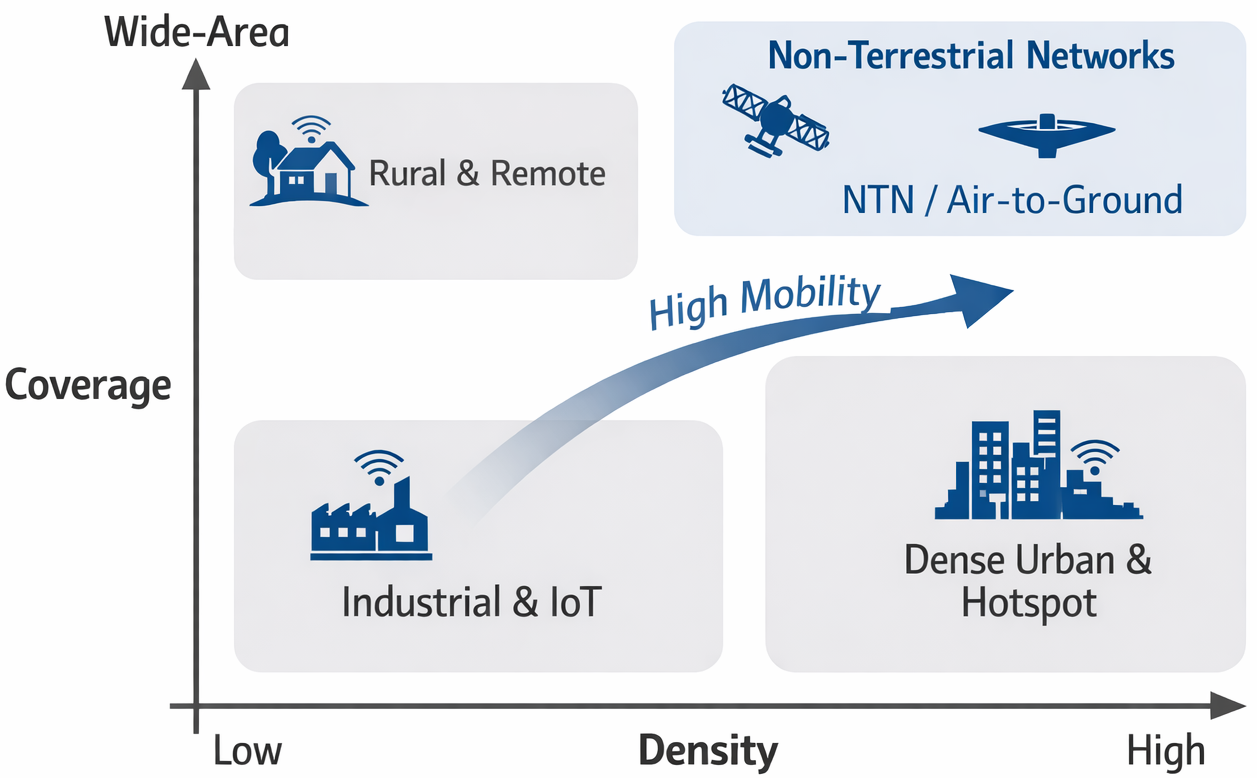}
    \caption{Unified taxonomy of 6G deployment scenarios based on density and coverage dimensions.}
    \label{fig:taxonomy}
\end{figure}

This taxonomy is illustrated in Fig.~\ref{fig:taxonomy}, which organizes 6G deployment scenarios according to two primary system dimensions: density and coverage. Additional aspects, such as mobility and infrastructure domains, are captured through annotations, highlighting their cross-cutting impact across different scenarios. As shown in the figure, dense urban and hotspot deployments are positioned in high-density regions, while rural and remote scenarios emphasize wide-area coverage. Industrial environments covered by Internet of Things (IoT) are typically characterized by localized coverage and moderate density, whereas NTNs or Air-to-Ground networks extend coverage beyond terrestrial limits.

Rather than categorizing scenarios by their nominal labels (e.g., urban, rural, or industrial), the proposed taxonomy is based on four key dimensions: density, coverage, mobility, and infrastructure domain. These dimensions capture the primary factors that influence radio access design and system behavior. In particular, mobility should not be interpreted as a single dominant attribute, as multiple mobility regimes may coexist within the same scenario (e.g., static, pedestrian, and vehicular users). Similarly, the infrastructure domain refers to the underlying connectivity layer, distinguishing between terrestrial and non-terrestrial components, such as satellite or high-altitude platforms, which introduce fundamentally different propagation and architectural constraints.

\subsection{Density-Driven Scenarios}

Density-driven scenarios feature a high concentration of users and devices within a limited geographic area. Typical examples include indoor hotspots, dense urban environments, and urban grid deployments. These scenarios impose stringent requirements on spectral efficiency, interference management, resource allocation, and support for heterogeneous traffic profiles. The coexistence of human-centric and machine-type communications further increases system complexity, requiring advanced scheduling and coordination mechanisms.

\subsection{Coverage-Driven Scenarios}

Coverage-driven scenarios prioritize wide-area connectivity over capacity, often in environments with low user density but large geographical extent. However, wide-area coverage does not necessarily translate into uniform or reliable service availability, particularly in challenging propagation conditions such as indoor or obstructed environments, which is especially relevant for non-terrestrial deployments. This category includes rural deployments, extreme long-distance coverage, and certain non-terrestrial use cases. 6G use cases are defined in \cite{giordani20206g}. Key challenges in these scenarios include maintaining reliable links under limited infrastructure, extending uplink coverage, and ensuring energy-efficient operation. These conditions emphasize the importance of link budget optimization, flexible spectrum usage, and infrastructure scalability.

\subsection{Mobility-Driven Scenarios}

Mobility-driven scenarios involve users or platforms moving at high speeds, such as high-speed trains, highway environments, and aerial platforms. In particular, satellite-based systems introduce very high relative velocities, while other aerial platforms such as high-altitude platform systems (HAPS) and unmanned aerial vehicles (UAVs) typically exhibit lower mobility regimes. These scenarios introduce challenges related to rapid channel variation, handover robustness, and synchronization. Ensuring service continuity and low latency under high mobility conditions requires enhancements in mobility management, predictive resource allocation, and robust waveform design.

\subsection{Infrastructure Domain: Terrestrial and Non-Terrestrial Integration}

A defining aspect of 6G scenarios is the inclusion of NTN, such as low Earth orbit (LEO), medium Earth orbit (MEO), geostationary (GEO) satellites, HAPS, and lower-altitude UAVs, spanning a wide range of orbital and atmospheric layers. \cite{10570308}. Unlike previous generations, in which NTN was treated as an extension, the Release 20 study considers the integration of terrestrial and non-terrestrial components a fundamental design aspect. This introduces new challenges related to propagation delay, intermittent connectivity, and mobility across heterogeneous domains, while also enabling global coverage and service continuity.

\subsection{Industrial and Deterministic Environments}

Industrial scenarios, such as indoor factory deployments, represent a distinct class characterized by stringent requirements on latency, reliability, and determinism. These environments often involve time-critical communications, precise synchronization, and tight integration with control systems. As a result, they require specialized support for deterministic networking, low-latency communication, and high reliability under controlled but demanding conditions.


The proposed taxonomy highlights that 6G deployment scenarios cannot be effectively understood in isolation. Instead, they are better interpreted as combinations of fundamental system dimensions, such as high density with high mobility, or wide coverage with non-terrestrial infrastructure. This multi-dimensional view enables a more flexible and scalable understanding of 6G requirements, and provides a foundation for identifying common design challenges across seemingly different scenarios.

In the next section, we build on this taxonomy to analyze how these scenarios translate into emerging service classes and capability demands.

\section{Emerging 6G Service Classes}
\label{sec:service}

Beyond deployment scenarios, the 3GPP TR 38.914 (Rel-20) study introduces a diverse set of services that significantly extend the scope of previous generations \cite{3gpp38914}. However, these services are typically presented as individual categories, which limits the ability to understand their combined impact on system design. To address this, we propose a unified classification of 6G services based on their fundamental operational characteristics.

\begin{figure} [!ht]
    \centering
    \includegraphics[width=0.85\linewidth]{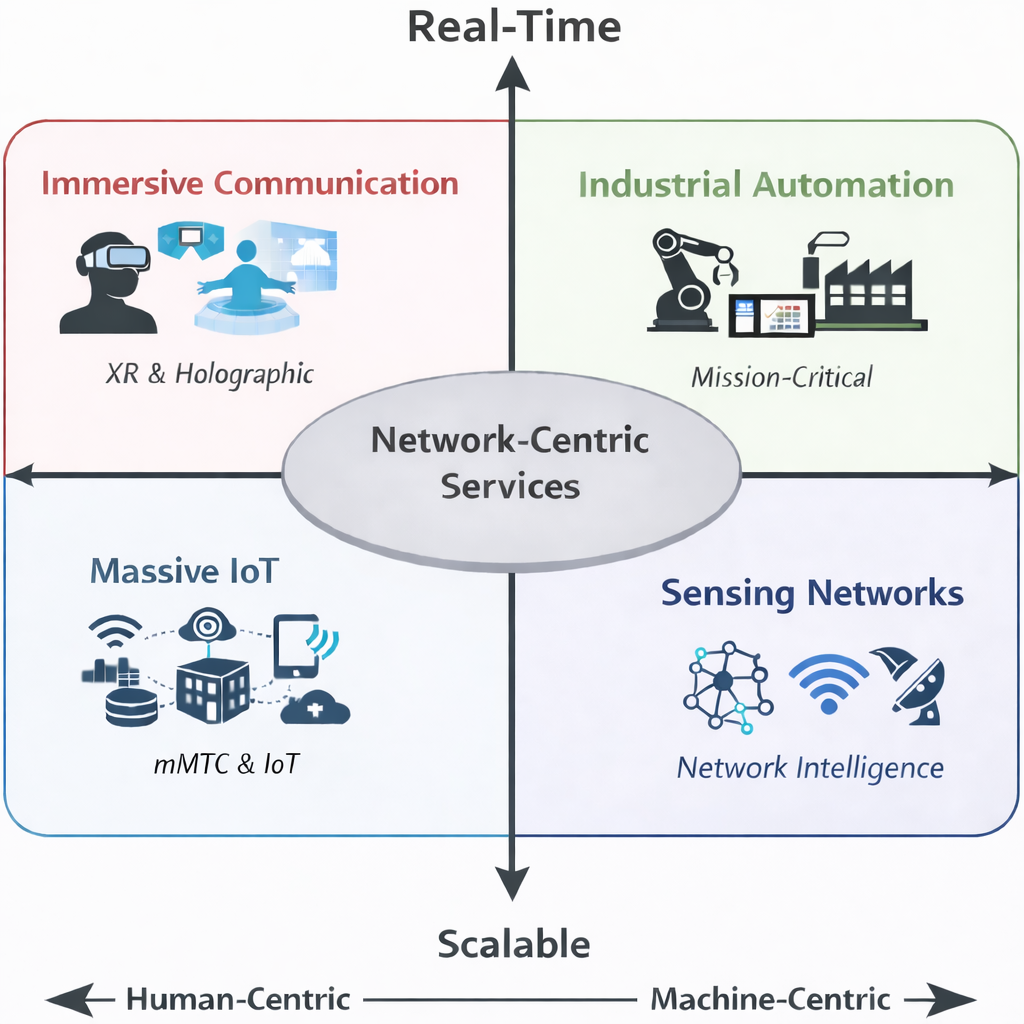}
    \caption{Unified classification of 6G service classes based on communication nature and system constraints.}
    \label{fig:services}
\end{figure}

Rather than following a service-by-service description, we organize 6G services along two key dimensions: the nature of the communication (human-centric vs. machine-centric) and the dominant system constraint (real-time vs. scalability). This abstraction enables a clearer understanding of how different services impose distinct and sometimes conflicting requirements on the network.

This classification is illustrated in Fig.~\ref{fig:services}, where 6G service classes are positioned along two key dimensions: the nature of communication (human-centric versus machine-centric) and the dominant system constraint (real-time versus scalability). This representation (Fig.~\ref{fig:services}) highlights the diversity of service requirements across 6G, ranging from real-time, human-centric applications to scalable, machine-centric deployments. It also emphasizes the emergence of network-centric capabilities, such as integrated sensing and AI-native functionalities, which extend beyond traditional service categories. However, it is important to note that AI is not confined to a specific service class, but rather acts as a cross-cutting enabler that can be applied across all service categories, reinforcing its role as a fundamental design enabler in 6G systems.

These service classes map onto the system requirement dimensions introduced in Section II, providing a structured basis to analyze how heterogeneous service demands translate into system-level constraints.

\begin{figure}[!ht]
    \centering
    \includegraphics[width=1\linewidth]{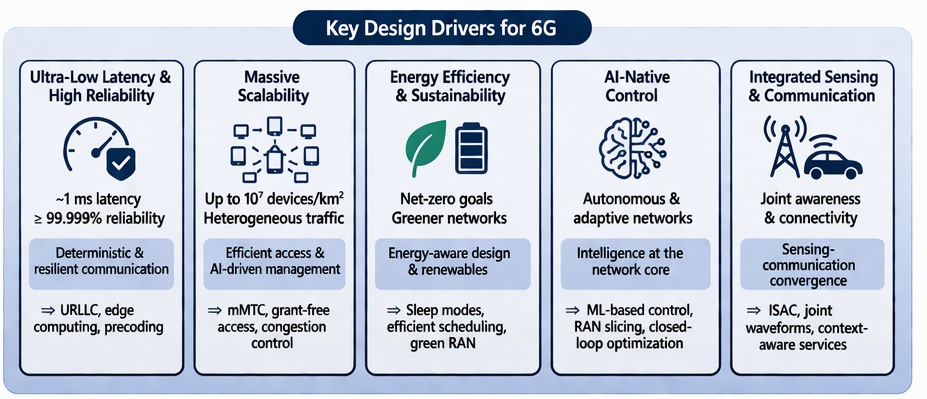}
    \caption{Key design drivers shaping 6G systems and their main characteristics.}
    \label{fig:drivers}
\end{figure}

\subsection{Human-Centric Real-Time Services}

This category includes immersive communication services such as extended reality (XR), holographic communications, and advanced multimedia applications. These services are characterized by strict latency and synchronization requirements, as well as high data rate demands. They require consistent quality of experience (QoE) and tight coordination across network layers, pushing the limits of radio performance and edge processing capabilities.

\subsection{Machine-Centric Massive Services}

Massive machine-type communication (mMTC) and large-scale IoT deployments fall into this category. These services prioritize scalability and energy efficiency over individual data rates. Supporting a large number of devices with sporadic traffic patterns requires efficient random-access mechanisms, lightweight signaling, and scalable resource management.

\subsection{Machine-Centric Real-Time Services}

Industrial automation, remote control, and mission-critical applications represent machine-centric services with strict real-time constraints. These use cases demand ultra-reliable low-latency communication (URLLC)-like performance, enhanced determinism, and precise synchronization. They also require strong integration with control systems and support for time-critical networking.

\subsection{Network-Centric and Cross-Domain Services}

A distinctive feature of 6G is the emergence of services that are not purely user-driven, but instead embedded within the network itself. These include integrated sensing, AI-native network functions, and context-aware services. Such capabilities blur the boundary between communication and computation, requiring the network to actively participate in data processing, inference, and environmental awareness.


The proposed classification highlights that 6G services span a wide spectrum of requirements, often combining real-time constraints with scalability demands. Unlike previous generations, where service categories could be addressed with relatively isolated solutions, 6G systems must support the coexistence of heterogeneous service types within a unified architecture.

This diversity of service requirements directly translates into a set of fundamental system constraints, including latency, reliability, scalability, energy efficiency, and computational capability. In the next section, we build on this classification to identify how these constraints converge into a set of key design drivers that shape the overall 6G system architecture.

\section{Key Design Drivers for 6G}
\label{sec:driver}
The analysis of deployment scenarios (Section~III) and service classes (Section~IV) reveals a set of fundamental system constraints that must be simultaneously addressed in 6G systems. Unlike previous generations, these constraints are not independent, but deeply interconnected, requiring coordinated design across multiple layers of the network architecture.

To capture this interplay, we identify a set of key design drivers that summarize the most critical and recurring requirements emerging from the 3GPP Rel-20 study. These drivers do not correspond to individual features, but rather to system-level principles that shape the design of 6G networks. 

Fig.~\ref{fig:drivers} provides an overview of the main design drivers emerging from this analysis, highlighting their core characteristics, while coverage-related aspects are further reflected in the architectural implications discussed in Section VI.

\subsection{Extreme Performance}

Extreme performance encompasses ultra-low latency, high data rates, and ultra-high reliability. This driver is essential to support immersive communications and mission-critical services, where end-to-end latency and service continuity are key requirements. Achieving this level of performance requires innovations in waveform design, edge computing, distributed architectures, and deterministic networking.

\subsection{Massive Scalability}

6G systems are expected to support an unprecedented number of devices, services, and data flows. Massive scalability extends beyond connectivity density to include the ability to manage heterogeneous traffic patterns and dynamic service requirements. This driver calls for efficient access mechanisms, scalable signaling, and AI-assisted resource management to ensure system stability at scale.

\subsection{Ubiquitous Coverage}

Ubiquitous coverage refers to the ability to provide seamless connectivity across diverse environments, including rural, urban, industrial, and non-terrestrial domains. This driver highlights the importance of integrating terrestrial and non-terrestrial networks into a unified system to enable global service availability and continuity.

\subsection{Native Intelligence}

AI is expected to become an intrinsic component of 6G systems \cite{11455199}, enabling adaptive, data-driven operation. Native intelligence supports functions such as network optimization, anomaly detection, traffic prediction, and resource allocation. In addition, it extends beyond traditional network functionalities to include distributed computing capabilities, enabling the execution of latency-sensitive tasks close to end users. This includes edge and in-network processing to support applications such as immersive services, real-time analytics, and AI inference. This driver implies a shift toward closed-loop, self-optimizing networks with distributed intelligence and computing resources across devices, edge, and core.

\subsection{Sustainability and Energy Efficiency}

Sustainability emerges as a fundamental requirement for 6G, driven by the need to reduce energy consumption and environmental impact. This includes energy-aware network design, dynamic resource adaptation, and the integration of renewable energy sources. Energy efficiency must be considered across all system components, from radio access to core network and devices.

\subsection{Integrated Sensing and Communication}

6G introduces the capability to jointly perform communication and sensing, enabling new applications such as environmental awareness, positioning, and context-aware services. This convergence requires new waveform designs, joint signal processing techniques, and cross-layer integration, thereby expanding the network's role beyond data transmission.

\subsection{Cross-Driver Interactions}

The identified design drivers are not isolated dimensions, but form a tightly coupled system of constraints and trade-offs. For instance, improving extreme performance may increase energy consumption, while AI-based optimization can help balance scalability and efficiency. As illustrated in Fig.~\ref{fig:drivers}, these drivers collectively define a multidimensional design space that 6G systems must navigate.

\begin{figure}
    \centering
    \includegraphics[width=1\linewidth]{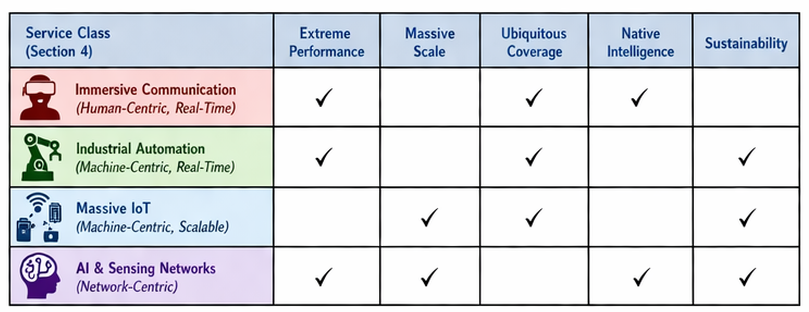}
    \caption{Mapping of 6G service classes to key design drivers}
    \label{fig:mapping}
\end{figure}

To further illustrate this relationship, Fig.~\ref{fig:mapping} maps the service classes introduced in Section~IV to the corresponding design drivers, highlighting how diverse service requirements converge into a common set of architectural imperatives.

\section{Implications for 6G System Design}
\label{sec:design}

The design drivers identified in the previous section provide a unified view of the fundamental constraints shaping 6G systems. Their impact lies in shaping concrete design choices across different layers of the network. In this section, we discuss the main implications of these drivers on 6G system architecture, radio design, and network operation.  

\begin{figure}
    \centering
    \includegraphics[width=1\linewidth]{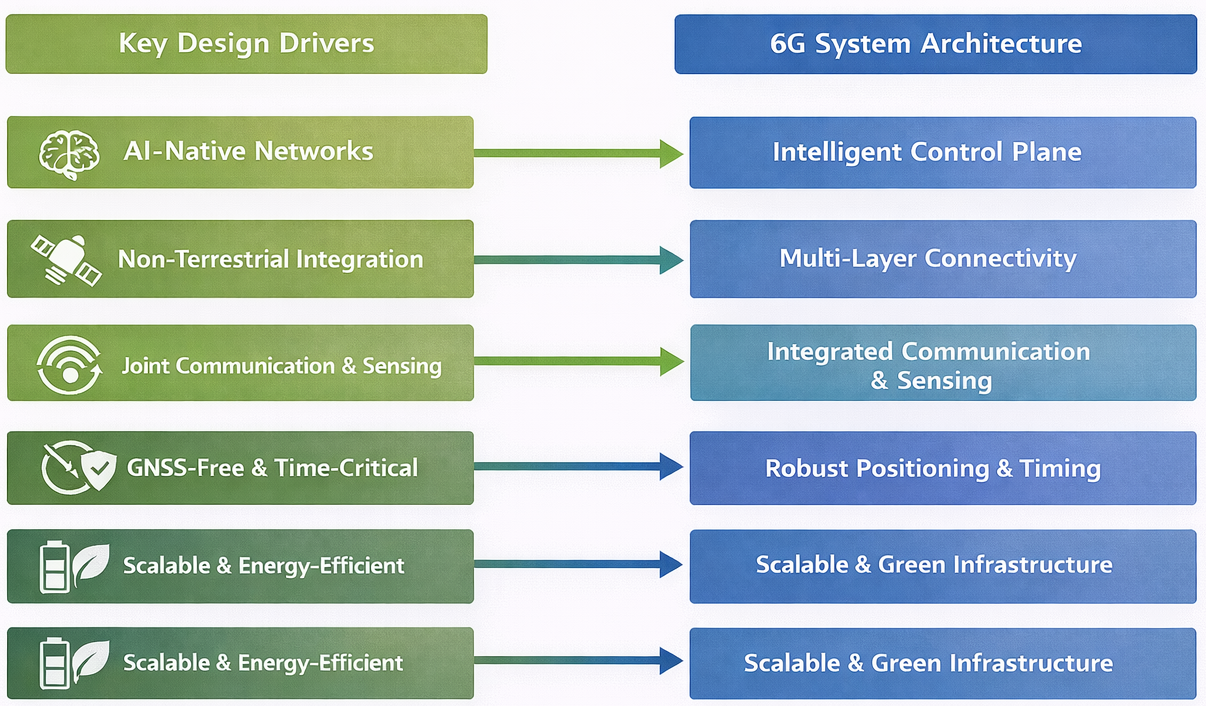}
    \caption{From design drivers to 6G system architecture.}
    \label{fig:architecture}
\end{figure}

This relationship is illustrated in Fig. \ref{fig:architecture}, which maps the key design drivers identified in Section V to their implications for 6G system architecture. The figure shows how these drivers translate into concrete architectural functionalities, reflecting a shift toward integrated, intelligent, and multi-domain systems.

\subsection{Architectural Evolution Toward Integrated and Distributed Systems}

The convergence of heterogeneous requirements, such as extreme performance, scalability, and ubiquitous coverage, necessitates a shift from monolithic architectures to more distributed and flexible system designs. In particular, integrating terrestrial and non-terrestrial networks requires a unified architecture that supports diverse propagation conditions and connectivity models.

This evolution is expected to promote tighter integration between access and core networks, as well as the distribution of functionalities across edge, cloud, and device layers. The resulting architecture must support seamless service continuity, dynamic resource allocation, and efficient coordination across domains.

\subsection{AI-Native Network Operation}

The inclusion of native intelligence as a core design driver implies that artificial intelligence will be embedded throughout the network stack \cite{mao2018deep}. Rather than being used as an auxiliary optimization tool, AI is expected to enable real-time adaptation to changing network conditions, traffic patterns, and service demands. In addition to network-centric optimization, AI will also support the processing of application-level data, including user traffic and sensing information, enabling in-network inference and real-time data analytics close to the end users.

This shift leads to the emergence of closed-loop control systems, where data collection, model inference, and decision-making are tightly integrated. It also introduces new requirements for data availability, model distribution, and trustworthiness, particularly in multi-domain and heterogeneous environments.

\subsection{Joint Communication and Sensing (JSC) Design}

The integration of sensing capabilities into communication systems represents a significant departure from traditional network design \cite{liu2022integrated}. This convergence requires the development of new waveform strategies, signal processing techniques, and resource allocation mechanisms that can simultaneously support data transmission and environmental sensing.

From a system perspective, this implies tighter coordination between physical and higher layers, as well as new trade-offs between communication performance and sensing accuracy. It also opens the door to new applications based on context awareness and environment perception.

\subsection{Support for GNSS-Free and Time-Critical Operation}

The requirement to operate without reliance on global navigation satellite systems (GNSS), particularly in non-terrestrial and constrained environments, introduces new challenges in positioning, synchronization, and timing distribution. At the same time, time-critical services demand highly deterministic communication and precise coordination across network elements.

These requirements motivate the development of alternative positioning techniques, enhanced synchronization mechanisms, and robust timing distribution over the air. They also reinforce the importance of reliability and resilience in 6G system design.

\subsection{Scalability and Energy-Aware Operation}

The combination of massive connectivity and sustainability goals requires 6G systems to operate efficiently at scale \cite{11358823}. This involves optimizing resource utilization, reducing signaling overhead, and enabling energy-aware operation across all network components.

Techniques such as dynamic resource allocation, sleep modes, and adaptive topology management will play a central role in achieving these objectives. Moreover, scalability and energy efficiency must be addressed jointly, as they are often interdependent and may introduce conflicting requirements.


The implications discussed above highlight that 6G system design is inherently multi-dimensional, requiring the simultaneous consideration of performance, scalability, intelligence, and sustainability. Rather than optimizing individual metrics in isolation, future systems must adopt a holistic design approach that balances these competing objectives.

This perspective reinforces the role of the identified design drivers as a unifying framework for guiding architectural and technological decisions in 6G. In the next section, we discuss the key challenges that remain open for future standardization and research.

\section{Open Challenges and Research Directions}
\label{sec:open}

While the 3GPP Rel-20 study provides a comprehensive foundation for 6G scenarios and requirements, and the design drivers identified in this work offer a structured interpretation of these elements, several challenges remain open for future standardization and research. These challenges arise from the intrinsic complexity of jointly addressing heterogeneous service demands, multi-domain deployments, and tightly coupled system constraints.

\subsection{Integration of Heterogeneous Network Domains}

One of the main challenges in 6G is the seamless integration of terrestrial and non-terrestrial networks into a unified architecture. This requires addressing fundamental differences in propagation conditions, latency, mobility patterns, and resource management across domains. Ensuring interoperability, efficient handover, and consistent quality of service in such heterogeneous environments remains an open issue.

\subsection{Scalable and Trustworthy AI-Native Systems}

The transition toward AI-native networks introduces challenges beyond scalability, including reliability, transparency, and trust \cite{11455199}. Deploying machine learning models across distributed network elements raises questions about data availability, model consistency, and robustness to dynamic conditions. Moreover, ensuring explainability and avoiding unintended behavior in critical applications are key concerns for future research.

\subsection{Joint Optimization of Communication and Sensing}

Although integrated sensing and communication offer significant potential, it also introduces new trade-offs between communication performance and sensing accuracy. Developing unified frameworks for jointly optimizing these functionalities across different network layers remains a complex challenge. This includes waveform design, resource allocation, and system-level coordination. Although integrated sensing and communication offer significant potential, it also introduces new trade-offs between communication performance and sensing accuracy. Developing unified frameworks for jointly optimizing these functionalities across different network layers remains a complex challenge. This includes waveform design, resource allocation, and system-level coordination. In addition, privacy and regulatory concerns, particularly in regions such as Europe, represent a major barrier to the adoption of sensing capabilities within communication systems.

\subsection{Energy Efficiency versus Performance Trade-offs}

Achieving extreme performance while maintaining energy efficiency represents a fundamental trade-off in 6G systems \cite{11359132}. High data rates, low latency, and large-scale connectivity often increase energy consumption, particularly in dense and heterogeneous deployments. Identifying mechanisms to balance these competing objectives, potentially through AI-driven optimization and adaptive resource management, is a key research direction.
\vspace{-4mm}
\subsection{GNSS-Free Operation and Synchronization}

Supporting operations without reliance on global navigation satellite systems introduces challenges in positioning, timing, and synchronization. Alternative approaches must be developed to provide accurate and reliable timing in diverse environments, including indoor, underground, and non-terrestrial scenarios. This is particularly critical for time-sensitive and mission-critical services.
\vspace{-4mm}
\subsection{System Complexity and Architectural Scalability}

The convergence of multiple design drivers increases system complexity, both in architecture and operation. Managing this complexity while maintaining scalability and efficiency is a major challenge for 6G systems. This includes the design of modular, flexible architectures and efficient orchestration mechanisms across network domains and layers.
\vspace{-4mm}
\subsection{Synthesis}

The challenges outlined above show that 6G design goes beyond performance, requiring the management of complexity, uncertainty, and trade-offs across multiple dimensions. Addressing these challenges will require coordinated efforts across standardization bodies, industry, and academia. Architectural trends such as network disaggregation, modular design, and open interfaces will be key to enabling flexible and interoperable systems. In particular, the evolution toward open, programmable, and cloud-native architectures—aligned with initiatives such as Open RAN—will support multi-vendor ecosystems, dynamic service deployment, and integration across heterogeneous terrestrial and non-terrestrial domains.

The proposed design-oriented framework provides a structured lens to address these challenges by linking scenarios, services, and requirements to fundamental system drivers, supporting the development of scalable and adaptable 6G systems across distributed computing and communication infrastructures.

Ultimately, the success of 6G will depend on translating these high-level requirements into practical and interoperable solutions capable of meeting the evolving demands of future wireless ecosystems.

\section{Conclusions}
\label{sec:con}
This article presented a design-oriented interpretation of the 3GPP Rel-20 study, linking 6G scenarios and requirements to a set of fundamental design drivers. By organizing these elements into a unified framework, we highlighted how diverse deployment conditions and service demands converge into common architectural principles.

The proposed perspective provides a structured basis for understanding early 6G standardization and supports the development of coherent, scalable system designs. As 6G evolves, such frameworks will be essential to bridge the gap between high-level requirements and practical implementation.

\bibliographystyle{IEEEtran}
\bibliography{references.bib}
\vspace{-1.1cm}
\begin{IEEEbiographynophoto}{Victor Monzon Baeza} 
(S'15, M'19, SM'24) received the B.Sc., M.Sc., and Ph.D. (Hons.) degrees in electrical engineering from the University Carlos III of Madrid, Spain, in 2013, 2015, and 2019, respectively. He received the Best Master's Thesis Award and the Best Ph.D. Awarded in 2014 and 2019, respectively. Since 2019, he has been a collaborator with Universitat Oberta de Catalunya. In 2022, he joined the SnT at the University of Luxembourg as a Research Associate. Currently, he is an Associate Professor at the University Internacional de La Rioja (UNIR). His research interests include satcom, 6G, NTN, IoT and AI-based communications.
\end{IEEEbiographynophoto}
\vspace{-1.1cm}
\begin{IEEEbiographynophoto}{Symeon Chatzinotas} (Fellow, IEEE) received the
M.Eng. degree in telecommunications from the Aristotle University of Thessaloniki, Greece, in 2003, and the M.Sc. and Ph.D. degrees in electronic
engineering from the University of Surrey, U.K., in 2006 and 2009, respectively. He is currently a Full Professor/Chief Scientist I and the Head of the Research Group SIGCOM, Interdisciplinary Centre for Security, Reliability and Trust, University of Luxembourg. In parallel, he is also an Adjunct Professor with the Department of Electronic Systems, Norwegian University of Science and Technology, an Eminent Scholar of Kyung Hee University, South Korea, and a Collaborating Scholar of the Institute of Informatics and Telecommunications, National Center for Scientific Research “Demokritos.” He has authored more than 800 technical papers in refereed international journals, conferences, and scientific books; and has received numerous awards and recognitions, including the IEEE Fellowship, the IEEE Distinguished Contributions Award, and the IEEE Harry Rowe Mimno Award. He has served in the editorial board of npj Wireless Technology, IEEE TRANSACTIONS ON COMMUNICATIONS, IEEE OPEN JOURNAL OF VEHICULAR TECHNOLOGY, and International Journal of Satellite Communications and Networking.
\end{IEEEbiographynophoto}

\end{document}